\documentclass{ws-rv9x6} 
\usepackage{hyperref}  

\hypersetup{
    colorlinks,
    citecolor=blue,
    filecolor=blue,
    linkcolor=blue,
    urlcolor=blue
}

\usepackage{ws-rv-har}    
\usepackage{ws-rv-thm}    
\usepackage{subfigure}    
\usepackage{ws-index}     
\usepackage{enumitem}
\usepackage[percent]{overpic}
\usepackage{xcolor}
\makeindex
\newindex{aindx}{adx}{and}{Author Index}       
\renewindex{default}{idx}{ind}{Subject Index}  

\begin{document}		
\newcommand{\ltsima}{$\; \buildrel < \over \sim \;$}
\newcommand{\lsim}{\lower.5ex\hbox{\ltsima}}
\newcommand{\gtsima}{$\; \buildrel > \over \sim \;$}
\newcommand{\gsim}{\lower.5ex\hbox{\gtsima}}
\newcommand{\bra}{\langle}
\newcommand{\ket}{\rangle}
\newcommand{\lprime}{\ell^\prime}
\newcommand{\lpp}{\ell^{\prime\prime}}
\newcommand{\mprime}{m^\prime}
\newcommand{\mpp}{m^{\prime\prime}}
\newcommand{\ci}{\mathrm{i}}
\newcommand{\dd}{\mathrm{d}}
\newcommand{\veck}{\mathbf{k}}
\newcommand{\vecx}{\mathbf{x}}
\newcommand{\vecr}{\mathbf{r}}
\newcommand{\vecv}{\mathbf{\upsilon}}
\newcommand{\vecw}{\mathbf{\omega}}
\newcommand{\vecj}{\mathbf{j}}
\newcommand{\vecq}{\mathbf{q}}
\newcommand{\vecl}{\mathbf{l}}
\newcommand{\vecn}{\mathbf{n}}
\newcommand{\lm}{\ell m}
\newcommand{\that}{\hat{\theta}}
\newcommand{\thatp}{\that^\prime}
\newcommand{\chip}{\chi^\prime}
\newcommand{\hs}{\hspace{1mm}}
\newcommand{\nar}{New Astronomy Reviews}
\def\gsim{~\rlap{$>$}{\lower 1.0ex\hbox{$\sim$}}}
\def\lsim{~\rlap{$<$}{\lower 1.0ex\hbox{$\sim$}}}
\def\Msun {\,\mathrm{M}_\odot}
\def\Jcrit {J_\mathrm{crit}}
\newcommand{\rsun}{R_{\odot}}
\newcommand{\mbh}{M_{\rm BH}}
\newcommand{\Msunyr}{M_\odot~{\rm yr}^{-1}}
\newcommand{\mdot}{\dot{M}_*}
\newcommand{\ledd}{L_{\rm Edd}}
\newcommand{\cmc}{{\rm cm}^{-3}}
\def\gsim{~\rlap{$>$}{\lower 1.0ex\hbox{$\sim$}}}
\def\lsim{~\rlap{$<$}{\lower 1.0ex\hbox{$\sim$}}}
\def\Msun {\,\mathrm{M}_\odot}
\def\Jcrit {J_\mathrm{crit}}

\def\simgreat{\lower2pt\hbox{$\buildrel {\scriptstyle >}
   \over {\scriptstyle\sim}$}}
\def\simless{\lower2pt\hbox{$\buildrel {\scriptstyle <}
   \over {\scriptstyle\sim}$}}
\def\msobh{M_\bullet^{\rm sBH}}
\def\zodot{\,{\rm Z}_\odot}
\newcommand{\lambdabar}{\mbox{\makebox[-0.5ex][l]{$\lambda$} \raisebox{0.7ex}[0pt][0pt]{--}}}

\def\na{NewA}%
\def\aj{AJ}%
\def\araa{ARA\&A}%
\def\apj{ApJ}%
\def\apjl{ApJ}%
\def\jcap{JCAP}

\def\pasa{PASA}

\def\apjs{ApJS}%
\def\ao{Appl.~Opt.}%
\def\apss{Ap\&SS}%
\def\aap{A\&A}%
\def\aapr{A\&A~Rev.}%
\def\aaps{A\&AS}%
\def\azh{AZh}%
\def\baas{BAAS}%
\def\jrasc{JRASC}%
\def\memras{MmRAS}%
\def\mnras{MNRAS}%
\def\pra{Phys.~Rev.~A}%
\def\prb{Phys.~Rev.~B}%
\def\prc{Phys.~Rev.~C}%
\def\prd{Phys.~Rev.~D}%
\def\pre{Phys.~Rev.~E}%
\def\prl{Phys.~Rev.~Lett.}%
\def\pasp{PASP}%
\def\pasj{PASJ}%
\def\qjras{QJRAS}%
\def\skytel{S\&T}%
\def\solphys{Sol.~Phys.}%

\def\sovast{Soviet~Ast.}%
\def\ssr{Space~Sci.~Rev.}%
\def\zap{ZAp}%
\def\nat{Nature}%
\def\iaucirc{IAU~Circ.}%
\def\aplett{Astrophys.~Lett.}%
\def\apspr{Astrophys.~Space~Phys.~Res.}%
\def\bain{Bull.~Astron.~Inst.~Netherlands}%
\def\fcp{Fund.~Cosmic~Phys.}%
\def\gca{Geochim.~Cosmochim.~Acta}%
\def\grl{Geophys.~Res.~Lett.}%
\def\jcp{J.~Chem.~Phys.}%
\def\jgr{J.~Geophys.~Res.}%
\def\jqsrt{J.~Quant.~Spec.~Radiat.~Transf.}%
\def\memsai{Mem.~Soc.~Astron.~Italiana}%
\def\nphysa{Nucl.~Phys.~A}%

\def\physrep{Phys.~Rep.}%
\def\physscr{Phys.~Scr}%
\def\planss{Planet.~Space~Sci.}%
\def\procspie{Proc.~SPIE}%

\newcommand{\rmp}{Rev. Mod. Phys.}
\newcommand{\ijmpd}{Int. J. Mod. Phys. D}
\newcommand{\sovjetp}{Soviet J. Exp. Theor. Phys.}
\newcommand{\jkas}{J. Korean. Ast. Soc.}
\newcommand{\PPVI}{Protostars and Planets VI}
\newcommand{\njp}{New J. Phys.}
\newcommand{\rap}{Res. Astro. Astrophys.}

\setcounter{chapter}{9}
\setcounter{page}{177}
\chapter[Growth and Feedback from the First Black Holes]{Growth and Feedback from the First Black Holes}
\label{wise}

\author[John~H. Wise]{John H. Wise}

\address{Center for Relativistic Astrophysics, School of Physics,   Georgia Institute of Technology, 837 State Street, Atlanta, GA   30332, USA\\ jwise@physics.gatech.edu}
\newcommand{\fesc}{\ifmmode{f_{\rm esc}}\else{$f_{\rm esc}$}\fi}
\newcommand{\fescs}{\ifmmode{f_{\rm esc}^\star}\else{$f_{\rm esc}^\star$}\fi}
\newcommand{\fgas}{\ifmmode{{f_{\rm gas}}}\else{$f_{\rm gas}$}\fi}
\newcommand{\cubecm}{\ifmmode{{\rm cm^{-3}}}\else{cm$^{-3}$}\fi}
\newcommand{\ztwo}{\ifmmode{{\rm [Z_2/H]}}\else{[Z$_2$/H]}\fi}
\newcommand{\zthree}{\ifmmode{{\rm [Z_3/H]}}\else{[Z$_3$/H]}\fi}
\newcommand{\li}{\noindent$\bullet$\quad}
\newcommand{\lcdm}{$\Lambda$CDM}
\newcommand{\flux}{erg s$^{-1}$ cm$^{-2}$ Hz$^{-1}$}
\newcommand{\emis}{erg s$^{-1}$ cm$^{-2}$ Hz$^{-1}$ sr$^{-1}$}
\newcommand{\sfr}{M$_\odot$ yr$^{-1}$ Mpc$^{-3}$}
\newcommand{\hsfr}{M$_\odot$ yr$^{-1}$}
\newcommand{\ssfr}{Gyr$^{-1}$}
\newcommand{\arp}{$a_{\rm rp}$}
\newcommand{\agrav}{$a_{\rm grav}$}
\newcommand{\eavg}{\ifmmode{\langle E_\gamma \rangle}\else{$\langle E_\gamma \rangle$}\fi}
\newcommand{\hst}{{\it HST}}
\newcommand{\jwst}{{\it JWST}}
\newcommand{\enzo}{{\sc enzo}}
\newcommand{\yt}{{\sc yt}}
\newcommand{\moray}{{\sc enzo+moray}}
\newcommand{\Ms}{\ifmmode{M_\odot}\else{$M_\odot$}\fi}
\newcommand{\vrms}{\ifmmode{v_{\rm rms}}\else{$v_{\rm rms}$}\fi}
\newcommand{\mmin}{M$_{min}$}
\newcommand{\hh}{H$_2$}
\newcommand{\Ol}{$\Omega_\Lambda$}
\newcommand{\Om}{$\Omega_M$}
\newcommand{\Ob}{$\Omega_b$}
\newcommand{\theat}{$t_{\rm{heat}}$}
\newcommand{\tcool}{$t_{\rm{cool}}$}
\newcommand{\rcool}{$r_{\rm{cool}}$}
\newcommand{\tcross}{$t_{\rm{cross}}$}
\newcommand{\tdyn}{$t_{\rm{dyn}}$}
\newcommand{\tkh}{$t_{\rm{KH}}$}
\newcommand{\tH}{$t_{\rm{H}}$}
\newcommand{\tvir}{\ifmmode{T_{\rm{vir}}}\else{$T_{\rm{vir}}$}\fi}
\newcommand{\mvir}{\ifmmode{M_{\rm{vir}}}\else{$M_{\rm{vir}}$}\fi}
\newcommand{\rvir}{\ifmmode{r_{\rm{vir}}}\else{$r_{\rm{vir}}$}\fi}
\newcommand{\rr}{$r_{200}$}
\newcommand{\jj}{\ifmmode{J_{21}}\else{$J_{21}$}\fi}
\newcommand{\flw}{\ifmmode{F_{LW}}\else{$F_{LW}$}\fi}
\newcommand{\kph}{\ifmmode{k_{\rm ph}}\else{$k_{\rm ph}$}\fi}
\newcommand{\tv}{$\langle T \rangle_{\rm v}$}
\newcommand{\tm}{$\langle T \rangle_{\rm m}$}
\newcommand{\msun}{{\rm\,M_\odot}} 
\newcommand{\lsun}{{\rm\,L_\odot}}
\newcommand{\zsun}{\ifmmode{\rm\,Z_\odot}\else{$\rm\,Z_\odot$}\fi}
\newcommand{\etal}{et al.\ }
\newcommand\tento[1]{$10^{#1}$}
\newcommand\halo[1]{\textit{Halo #1}.--}
\newcommand{\hi}{H {\sc i}}
\newcommand{\hii}{H {\sc ii}}
\newcommand{\hei}{He {\sc i}}
\newcommand{\heii}{He {\sc ii}}
\newcommand{\heiii}{He {\sc iii}}
\newcommand{\msigma}{$M$-$\sigma$}
\newcommand{\kms}{\ifmmode{\rm\,km\,s^{-1}}\else{km s$^{-1}$}\fi}

\newcommand{\papers}[1]{\begin{quote}{\color{red}{\bf Papers:} #1}\end{quote}}

\newcommand\mysec[1]{\noindent\textbf{#1}.---}
\newcommand\unit[1]{\; \textrm{#1}}
\def\hr{\begin{center}
  \line(1,0){250}
\end{center}
}
\newcommand\refs{({\bf Refs?})}

\def\newacronym#1#2#3{\gdef#1{#3 (#2)\gdef#1{#2}}}
\newacronym{\EM}{EM}{electromagnetic}
\newacronym{\UV}{UV}{ultraviolet}
\newacronym{\dm}{DM}{dark matter}
\newacronym{\ism}{ISM}{interstellar medium}
\newacronym{\igm}{IGM}{intergalactic medium}
\newacronym{\sfr}{SFR}{star formation rate}
\newacronym{\sfh}{SFH}{star formation history}
\newacronym{\amr}{AMR}{adaptive mesh refinement}
\newacronym{\sdsc}{SDSC}{San Diego Supercomputer Center}
\newacronym{\sdss}{SDSS}{Sloan Digital Sky Survey}
\newacronym{\cra}{CRA}{Center for Relativistic Astrophysics}
\newacronym{\nr}{NR}{numerical relativity}
\newacronym{\ornl}{ORNL}{Oak Ridge National Laboratory}
\newacronym{\jwst}{JWST}{James Webb Space Telescope}
\newacronym{\hst}{HST}{Hubble Space Telescope}
\newacronym{\alma}{ALMA}{Atacama Large Millimeter Array}
\newacronym{\lisa}{LISA}{Laser Interferometer Space Antenna}
\newacronym{\ligo}{LIGO}{Laser Interferometer Gravitational Wave Observatory}
\newacronym{\sph}{SPH}{smooth particle hydrodynamics}
\newacronym{\tsi}{TSI}{Terascale Supernova Initiative}
\newacronym{\wmap}{WMAP}{Wilkinson Microwave Anisotropy Probe}
\newacronym{\cmb}{CMB}{cosmic microwave background}
\newacronym{\ibbh}{IBBH}{intermediate binary black hole}
\newacronym{\grb}{GRB}{Gamma-ray burst}
\newacronym{\cmb}{CMB}{cosmic microwave background}
\newacronym{\mhd}{MHD}{magneto-hydrodynamic}
\newacronym{\mw}{MW}{Milky Way}
\newacronym{\dm}{DM}{dark matter}
\newacronym{\imf}{IMF}{initial mass function}
\newacronym{\sfr}{SFR}{star formation rate}
\newacronym{\sfh}{SFH}{star formation histories}
\newacronym{\sdsc}{SDSC}{San Diego Supercomputer Center}
\newacronym{\sdss}{SDSS}{Sloan Digital Sky Survey}
\newacronym{\cra}{CRA}{Center for Relativistic Astrophysics}
\newacronym{\nr}{NR}{numerical relativity}
\newacronym{\ornl}{ORNL}{Oak Ridge National Laboratory}
\newacronym{\jwst}{JWST}{James Webb Space Telescope}
\newacronym{\hst}{HST}{Hubble Space Telescope}
\newacronym{\alma}{ALMA}{Atacama Large Millimeter Array}
\newacronym{\lisa}{LISA}{Laser Interferometer Space Antenna}
\newacronym{\ligo}{LIGO}{Laser Interferometer Gravitational Wave Observatory}
\newacronym{\sph}{SPH}{smooth particle hydrodynamics}
\newacronym{\tsi}{TSI}{Terascale Supernova Initiative}
\newacronym{\wmap}{WMAP}{Wilkinson Microwave Anisotropy Probe}
\newacronym{\cmb}{CMB}{cosmic microwave background}
\newacronym{\ibbh}{IBBH}{intermediate binary black hole}
\newacronym{\grb}{GRB}{gamma-ray burst}
\newacronym{\mhd}{MHD}{magneto-hydrodynamics}
\newacronym{\agn}{AGN}{active galactic nuclei}
\newacronym{\hpc}{HPC}{High-performance Computing}
\newacronym{\imf}{IMF}{initial mass function}
\newacronym{\fld}{FLD}{flux limited diffusion}
\newacronym{\bssn}{BSSN}{Baumgarte-Shapiro-Shibata-Nakamura}
\newacronym{\imbh}{IMBH}{intermediate-mass black hole}
\newacronym{\mbh}{MBH}{massive black hole}
\newacronym{\vet}{VET}{variable Eddington tensor}
\newacronym{\ska}{SKA}{Square Kilometer Array}

\def\gpu#1{graphics processing unit#1 (GPU#1)\gdef\gpu{GPU}}
\def\lf#1{luminosity function#1 (LF#1)\gdef\lf{LF}}
\def\bbh#1{binary black hole#1 (BBH#1)\gdef\bbh{BBH}}
\def\smbh#1{supermassive black hole#1 (SMBH#1)\gdef\smbh{SMBH}}
\def\bh#1{black hole#1 (BH#1)\gdef\bh{BH}}
\def\gc#1{globular cluster#1 (GC#1)\gdef\gc{GC}}
\def\gcmf#1{globular cluster mass function#1 (GCMF#1)\gdef\gcmf{GCMF}}
\def\mdf#1{metallicity distribution function#1 (MDF#1)\gdef\mdf{MDF}}
\def\gmc#1{giant molecular clouds#1 (GMC#1)\gdef\gmc{GMC}}
\def\ism#1{interstellar medium#1 (ISM#1)\gdef\ism{ISM}}
\def\igm#1{intergalactic medium#1 (IGM#1)\gdef\igm{IGM}}
\def\sn#1{supernova#1 (SN#1)\gdef\sn{SN}}
\def\imf#1{initial mass function#1 (IMF#1)\gdef\imf{IMF}}
\def\sfr#1{star formation rate#1 (SFR#1)\gdef\sfr{SFR}}
\def\sfh#1{star formation history#1 (SFH#1)\gdef\sfh{SFH}}


\newenvironment{lquote}{%
  \list{}{%
    \rightmargin0pt}%
    \item\relax
  }
{\endlist}

\let\oldenumerate=\enumerate
\let\endoldenumerate=\endenumerate
\renewenvironment{enumerate}{%
  \begin{oldenumerate}%
    \setlength{\itemsep}{0.0ex}%
    \setlength{\parskip}{0.1ex}%
  }%
  {%
  \end{oldenumerate}%
}

\let\olditemize=\itemize
\let\endolditemize=\enditemize
\renewenvironment{itemize}{%
  \begin{olditemize}%
    \setlength{\itemsep}{0.0ex}%
    \setlength{\parskip}{0.25ex}%
  }%
  {%
  \end{olditemize}%
}

\let\olddescription=\description
\let\endolddescription=\enddescription
\renewenvironment{description}{%
  \begin{olddescription}%
    \setlength{\itemsep}{0.0ex}%
    \setlength{\parskip}{0.25ex}%
  }%
  {%
  \end{olddescription}%
}

\footnotetext{Preprint~of~a~review volume chapter to be published in Latif, M., \& Schleicher, D.R.G., ``Growth and  Feedback from the First Black Holes'', Formation of the First Black Holes, 2018. \quad \textcopyright Copyright World Scientific Publishing Company, \url{https://www.worldscientific.com/worldscibooks/10.1142/10652}}

\begin{abstract}
  Regardless of their initial seed mass, any active galactic nuclei  observed at redshifts $z \ge 6$ must have grown by several orders of  magnitude from their seeds.  In this chapter, we will discuss the physical processes and latest research on black hole growth and  associated feedback after seed formation.  Fueling is initially slowed down by radiative feedback from the black hole itself and supernova explosions from nearby stars.  Its growth however accelerates once the host  galaxy grows past a critical mass.
\end{abstract}


\section{Observational Motivation}

Since the beginning of the century, there have been approximately 40 supermassive black holes (SMBHs) discovered above a redshift of six when the universe was barely one billion years old \citep{Fan2003,Willot2010, Banados2014}.  The vast majority of these distant quasars exist in the bright-end of the active galactic nuclei (AGN) luminosity function \citep{Kashikawa2015}.  The most distant observed SMBH has a redshift of $z = 7.54$ and a mass of $\sim 8 \times 10^8~\Ms$
\citep{Banados18} while the most massive SMBH observed in the early Universe has a mass of $\sim 1.2 \times 10^{10}~\Ms$ at $z = 6.30$ \citep{Wu15}.  These rare luminous objects are only the tip of the iceberg of the high-redshift SMBH population, only occurring about once per comoving Gpc$^3$ \citep{Fan06_Review}, with the remaining population having more moderate growth histories that exist in more normal galaxies like our Milky Way \citep{Gultekin09}.

In addition to inspecting growth histories of individual SMBHs, cumulative statistics and quantities, such as the total BH mass density and AGN/galaxy luminosity functions, place further constraints on BH growth in the early Universe.  Because only the brightest AGN are detected at $z \ge 6$, the Soltan argument \citep{Soltan82} can estimate the BH mass density from the AGN luminosity function.  It argues that accreted mass density into SMBHs, which is derived from the integrated quasar light and some radiative efficiency factor, must be less than the total BH mass density, i.e. $\rho_{\rm acc} \le \rho_{\rm BH}$.  Another important statistical measure is the present-day \msigma{} relation that suggests that there is some connection between galaxy and BH growth \citep{Silk98, Ferrarese00, Gebhardt00, Gultekin09}.  Its behavior at high-redshift is difficult to observe, but in theoretical models, it provides a window in how the \msigma{} relation is established as both the host galaxy and central BH grow over billions of years \citep{Robertson06, Volonteri09}.  By combining the \msigma{} relation and galaxy luminosity (or mass) function, the total BH mass density in SMBHs with $M_{\rm BH} > 10^6~\Ms$ rises from $\sim 10^{-3}~\Ms$ per comoving Mpc$^{-3}$ at $z = 7$ to the present-day value of $5 \times 10^5~\Ms$ per comoving Mpc$^{-3}$ \citep{Shankar04, Somerville08, Willot2010, Schulze11, Volonteri17}. A more detailed account on the current status of observations is given in Chapter 12.

These individual and statistical observational SMBH measures provide a basis for the theoretical models of early BH growth to match at $z \sim 6$.  This chapter will explore the initial buildup of the first black holes and its impact on the first galaxies, following its behavior prior to galaxy formation to the end of reionization when the farthest quasars have been observed.  First, we will estimate the growth rates from abundance matching in \S~\ref{sec:estimate}.  Then in \S~\ref{sec:processes}, we will review the important physical processes that govern the gas inflow and associated feedback. \S~\ref{sec:initial} covers the bulk of this chapter and focuses on the environment and growth shortly after the formation of the first seed black holes (see Chapters 6 and 7 for their formation scenarios).  Then we follow their journey into larger halos in \S~\ref{sec:larger}, becoming central massive black holes in the first generations of galaxies.  We close this topic with a discussion of
outstanding questions and the possible paths forward to illuminate these mysteries in \S~\ref{sec:future}.

\begin{table}[htp]
\caption{default}
\begin{center}
\begin{tabular}{|c|c|}
Abbreviation & full name\\ \hline
BH & black hole\\
SMBH & supermassive black hole\\
DCBH & direct collapse black hole\\
AGN & active galactic nuclei\\
IGM & intergalactic medium\\
ISM & interstellar medium
\end{tabular}
\end{center}
\label{abrsuper}
\end{table}%

\section{Host Halo Progenitors and Timescale Estimates} 
\label{sec:estimate}

\begin{figure}[t]
\centering
\includegraphics[width=0.49\textwidth]{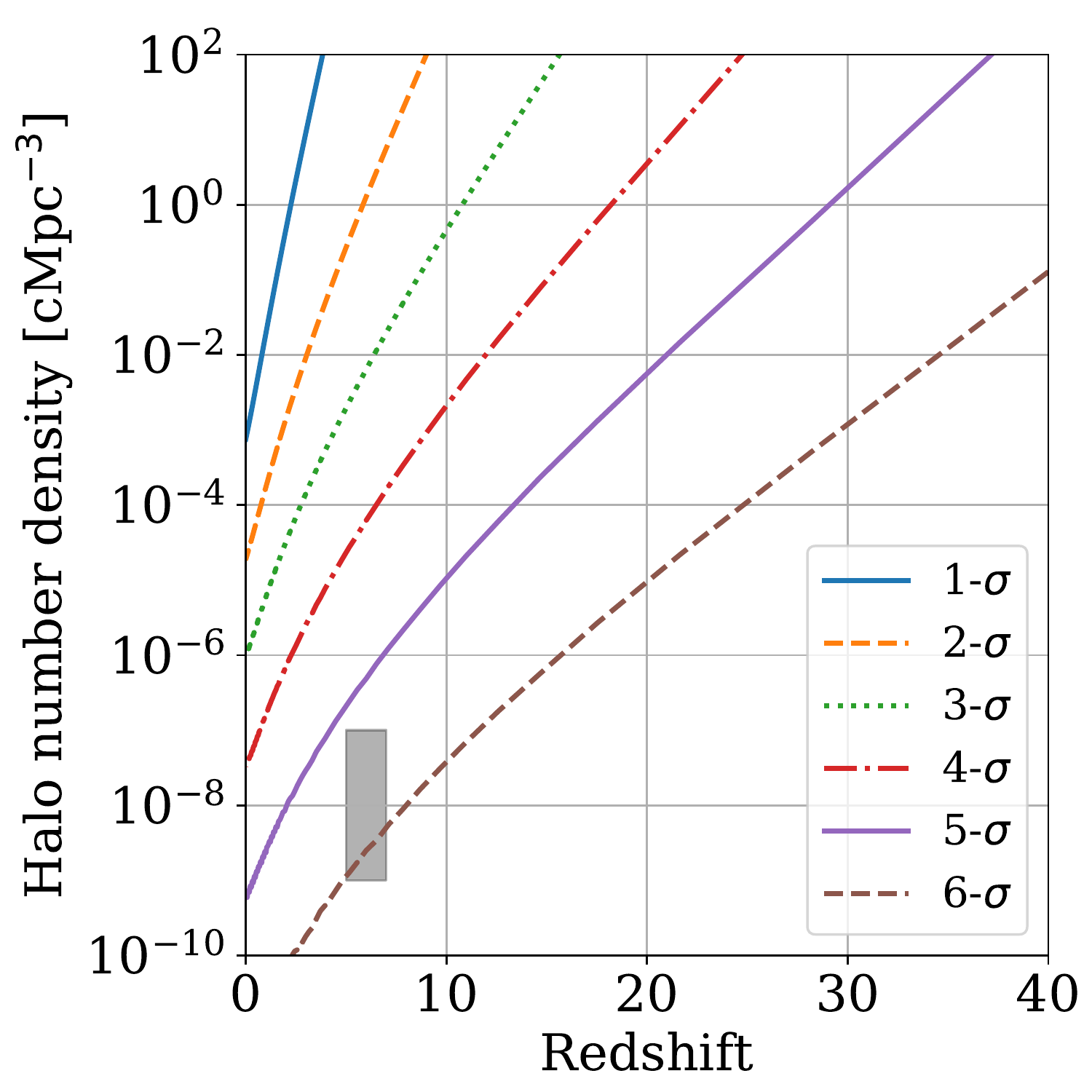}
\hfill
\includegraphics[width=0.49\textwidth]{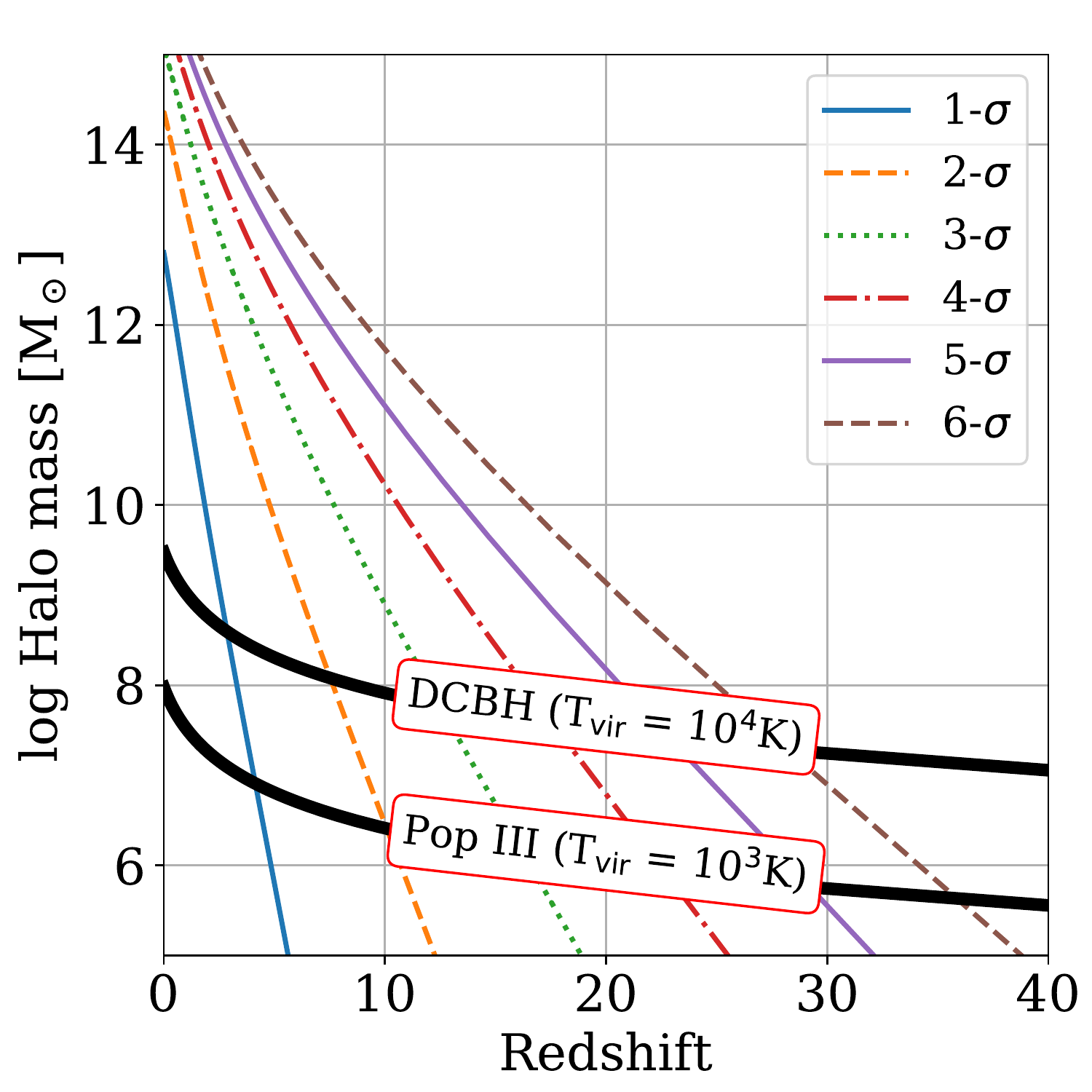}
\caption{\label{fig:ps} {\it Left panel:} Each line shows the comoving number density of dark matter halos for a particular rarity, i.e. $n$-$\sigma$.  The shaded box shows an estimate of the SMBH number density at $z \sim 6$, considering a duty cycle between 1--100\%.  Simple abundance matching of $z \sim 6$ AGN suggests that their host halos are very rare density fluctuations (5--6 $\sigma$) in large scale structure. {\it Right panel:} Dark matter halo masses for particular $\sigma$ values compared with critical halo masses for Population III (metal-free) star formation and direct-collapse black hole (DCBH) formation.  The 5- and 6-$\sigma$ tracks can be used to estimate the $z \sim 6$ progenitor halo masses at earlier redshifts, suggesting that they hosted their first star around $z \sim 35$ (80 million years after the Big Bang) and started galaxy formation around $z \sim 25$ (130 million years after the Big Bang).}
\end{figure}

To begin, we can use the comoving number density of the brightest $z \sim 6$ AGN to estimate the typical seed formation time in the latest Planck cosmology \citep{Planck2016}.  Using an ellipsoidal variant of the Press-Schechter formalism \citep{Press1974, smt01}, we calculate the number density of dark matter halos, shown in the left panel of Figure \ref{fig:ps}.  Here $n$-$\sigma$ denotes the rarity of each halo in a Gaussian density field, and the grey shaded area shows the estimated number density, assuming that some fraction $f = 0.01-1$ of SMBHs are currently in an active phase \citep{Luo11}.  At $z = 6$, this simple abundance matching exercise suggests that the brightest AGN are hosted in very rare and massive halos (5-$\sigma$ to 6-$\sigma$).  These halos have masses around $10^{13}~\Ms$ at this redshift, shown in the right panel of Figure \ref{fig:ps}.  We can then trace back the most massive progenitor mass at the same $\sigma$ value, where it must have formed its first star at redshifts $z \simeq 30-35$ and could efficiently cool through hydrogen atomic line cooling at redshifts $z \simeq 25-30$.  At these redshifts, the universe is around 100 million years old.  Thus, the seed black holes only have 800 million years to grow by 4--7 orders of magnitude in order to transform into the observed SMBHs at redshift 6.  Aiding this difficult ascent, their host halos will grow by similar factors.

\section{Relevant Physical Processes and Methods}
\label{sec:processes}

The growing BH will have to compete with star formation and feedback processes for any accreted gas.  Through a myriad of mergers and smooth IGM (intergalactic medium) accretion, the halo will have an ample gas supply, but the big question is what fraction makes the long journey to the SMBH event horizon, affected by many physical processes along the way.

\subsection{Accretion}
\label{sec:accretion}

Any gas inside of the innermost stable orbit at three Schwarzschild radii ($R_{\rm sch} = 2GM_{\rm BH}/c^2$) will be accreted by the BH. However, it first must overcome the ``angular momentum barrier'' to exist at such a small radius.  The gas inside pre-galactic halos is mainly turbulent from virialization \citep{Wise2007, Greif2008} and has some coherent rotational motions that are usually around 30--50\% of the Keplerian value $\sqrt{GM/R}$ \citep{Regan09}.  Being a turbulent medium, there always exists some gas with very low specific angular momentum \citep{Wise2008} that can migrate toward the central regions without being rotationally supported \citep{Lodato2006,Lodato2007}.

\begin{figure}[t]
\centering
\includegraphics[width=\textwidth]{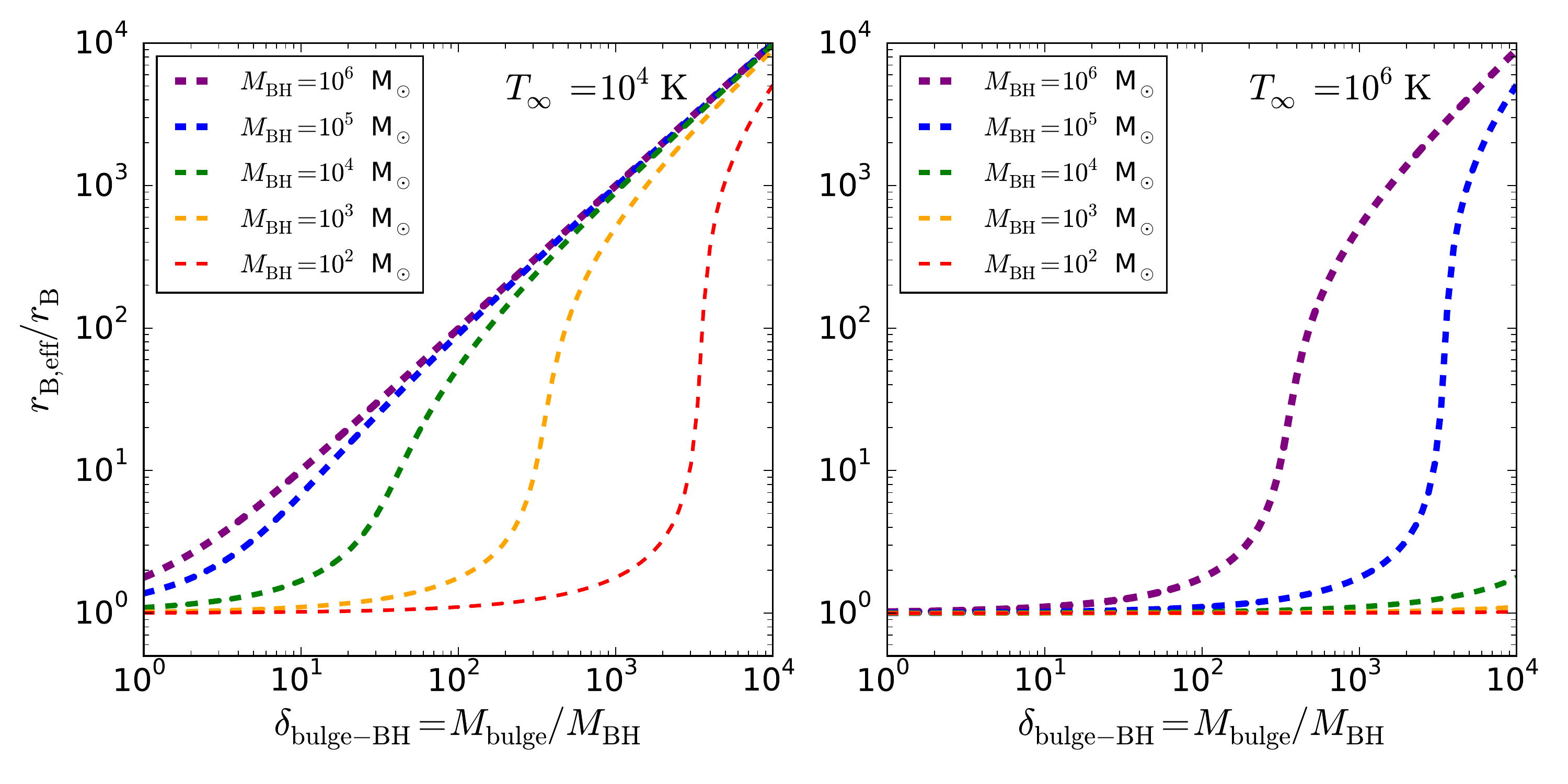}
\caption{\label{fig:bondi} The effective Bondi radius in units of the Bondi radius as a function of bulge-to-BH mass ratio for BH masses ranging from $10^2$ to $10^6~\Ms$ that reside in gas reservoir with $T = 10^4 \unit{K}$ (left) and $10^6 \unit{K}$ (right).  Adopted from  \citet{Park2016},  \textcopyright AAS. Reproduced with permission.}
\end{figure}

Once the gas is within the central regions of the halo, the next barrier is becoming gravitationally bound to the BH, which was first described by \citet{Bondi44} and \citet{Bondi52} in the context of point masses and a spherically symmetric system.  For gas with an adiabatic equation of state ($\gamma = 5/3$), this transition occurs at the Bondi radius, $r_{\rm B} \equiv GM_{\rm BH}/c_{\rm s}^2$, where $c_{\rm s}$ is the sound speed.  The gas entering this sphere is assumed to accrete onto the BH at the Bondi-Hoyle rate, $\dot{M}_{\rm BH} = 4\pi G^2 M_{\rm BH}^2 \rho / (c_{\rm s}^2 + v_{\rm rel}^2)^{3/2}$, where $\rho$ is the gas density at the Bondi radius and $v_{\rm vel}$ is the gas velocity relative to the BH. \citet{Park2016} modified this prescription to include an external gravitational potential.  The most likely sources for this potential are a stellar bulge or a cusped dark matter halo.  They found that the
effective Bondi radius, shown in Figure \ref{fig:bondi}, dramatically increases when the total bulge mass exceeds $10^6~\Ms$ that would boost BH fueling.

Once the gas flows into the Bondi radius, it is assumed that some fraction will deposit onto the accretion disk, in which it will lose further angular momentum through viscous forces \citep{Power11} and magnetic instabilities \citep{Balbus91}.  It is a common assumption that the accretion rate onto the BH is equal to the Bondi-Hoyle rate or some scaled version \citep{Schaye10}, limited by the Eddington rate, $\dot{M}_{\rm Edd} = 4\pi GM_{\rm BH} m_{\rm p}/ \epsilon c \sigma_{\rm T}$, where $m_{\rm p}$ and $\sigma_{\rm T}$ are the proton mass and Thomson scattering cross-section.  Here $\epsilon$ is the radiative efficiency factor that dictates what mass-energy fraction of the accreted material is converted into radiation.  If the BH grows at the Eddington limit, the $e$-folding time is $4.5 \times 10^7 (\epsilon/0.1)^{-1} \unit{yr}$.  Applying this growth rate to the early universe, the largest SMBH at redshift 6 only has 15--18 $e$-folding times, corresponding to a mass ratio of $\log(M_{\rm BH}/M_0) = 6.5 - 7.8$, to grow to $10^9~\Ms$, where $M_0$ is the seed BH mass.  In the following subsection, we will see that the associated feedback can hamper SMBHs from growing at this limit.

On the other hand, the Eddington limit is derived in spherical symmetry.  Super-Eddington accretion rates can occur when
\begin{enumerate}
\item the medium is not spherically symmetric or is porous, such as a  disk \citep{Abramowicz88, Novak12} or (supersonic)  turbulence \citep{VanBorm2013, VanBorm2014}, or 
\item the radiation can be trapped with the gas flow, is advected with the fluid flow, and does not impart all of its momentum to the surrounding medium \citep{Begelman79, Jiang14}.
\end{enumerate}
In these cases, BH accretion rates can easily exceed the Eddington rate up to a factor of $\sim$200 for short periods at a radiative efficiency of $\sim$5\% \citep{Jiang14}.  These bursts somewhat alleviate the constraints on the growth history of the most massive SMBHs at $z \gsim 6$ \citep{Volonteri2005}.

\subsection{Feedback}
\label{sec:feedback}

A fraction of the accreted gas mass-energy will inevitably be returned to the surrounding environment \citep{Fabian12}.  What fraction is unclear, but given the tremendous growth rates necessary to build the observed SMBHs, any fraction may have a substantial effect on the subsequent fueling.  These feedback processes can be broadly categorized into radiative and mechanical feedback, both of which can launch winds from AGN that are ubiquitous in intrinsic quasar absorption spectra \citep{Hamann08, Grier15}.  They will transition between a radiative mode (quasar mode) when $\dot{M}_{\rm  BH} \lsim 0.02\,\dot{M}_{\rm Edd}$ and a mechanical mode (radio mode) at higher accretion rates \citep{Croton06, McNamara07}.

Many different astrophysical processes, such as blackbody radiation and inverse Compton scattering in the disk and its corona respectively, can be categorized into radiative feedback with the luminosity $L = \epsilon c^2 \dot{M}_{\rm BH}$, which is appropriate when radiation is decoupled from the gas.  However in a dense optically-thick environment, such as the disk or jet, the radiation is trapped as photons scatter within the medium, causing $L \propto
\ln\dot{M}_{\rm BH}$, that limits the amount of feedback \citep{Begelman79}.  This radiation will photo-heat the surrounding gas, creating a \hii{} region, and will impart its momentum $|p| = E/c$ onto the baryons, i.e. radiation pressure.  Heating the gas will reduce its susceptibility to accretion, correspondingly reducing the Bondi-Hoyle accretion rate \citep{Alvarez2009, Park12, Jeon12}. In addition, the radiation pressure will slow  down or, in some cases, reverse the gas inflow toward the BH \citep{Cowie78,  Pacucci2015, Kohei}.

Observations have shown that a significant portion of energy is released through bipolar relativistic jets \citep{Blandford77, Bridle84, Begelman84, Pounds03, Peterson06}, also known as mechanical feedback, that entrains some fraction of mass along its journey through the host galaxy.  These jets can inflate cavities in the largest of AGN in galaxy clusters \citep{Fabian02, McNamara05}. Contrary to radiative feedback, the kinetic energy of the jet will be
lost as it exits the SMBH gravitational potential well.  As the jet propagates through the immediate region \citep{McKinney06, Sadowski13} and the interstellar medium (ISM), it will shock-heat the gas, eventually launching an outflow from the host halo/galaxy \citep{Sijacki2007, Dubois12, Bieri17}.

\subsection{Methods}
\label{sec:methods}

Capturing the growth and feedback of SMBHs in a cosmological setting is a difficult task primarily originating from the enormous dynamic spatial and temporal range of the problem.  To follow the full growth history, it is essential to calculate the host halo merger tree, as the mergers bring in new gas supply for further accretion and other SMBHs that will merge after some time.  Furthermore, following the host galaxy evolution is critical because there will be a feedback loop between the two \citep{Silk98, Hopkins06, Somerville15}.  There are two major players in modeling galaxies and BHs over their lifetimes---semi-analytical models and numerical simulations---both of which can include the physics previously described directly or indirectly.  In the process, the physical properties of the SMBH and galaxy are converted into key observables \citep{Illustris, EAGLE} and compared and tuned against current observations, which allows for predicting observations with the next-generation observatories \citep{Barrow17}.  Typical observables include luminosity functions, event rates for gravitational wave detections, mass functions, number counts on the sky as a function of redshift, and their spectra. 

\paragraph{Semi-analytical models} use a halo merger tree as a basis for galaxy and BH formation and evolution \citep{Kauffmann93,Somerville08, Benson10}.  It can be calculated either from extended Press-Schetcher \citep{Bond91, Lacey93}, perturbative Lagrangian techniques \citep{Monaco02, Munari17}, or extracted from an $N$-body simulation \citep{Guo14}.  Physically motivated models are included into each halo and followed through the merger tree.  These recipes control processes such as gas cooling, star formation, chemical enrichment, BH accretion, stellar and gas radial distributions, and any associated feedback mechanisms.  Specific to
the BH, different accretion models (e.g. Bondi, Eddington-limited, super-Eddington), feedback models (e.g. radiative, jets, mass-loading, efficiency factors), accretion disk properties (e.g. slim disk), and what triggers bursts of high accretion (e.g. mergers) can be altered to inspect the resulting AGN and galaxy evolutionary track to understand how each process governs their properties and of their importance.  The most serious disadvantage of semi-analytical models is the lack of spatial information, such as clustering, but they are computationally inexpensive and can be used to explore vast regions of parameter space.  Furthermore, they can be utilized in Markov Chain Monte Carlo methods to place constraints on the importance of each physical process \citep{Lu11}.

\begin{figure}[t]
\centering
\includegraphics[width=\textwidth]{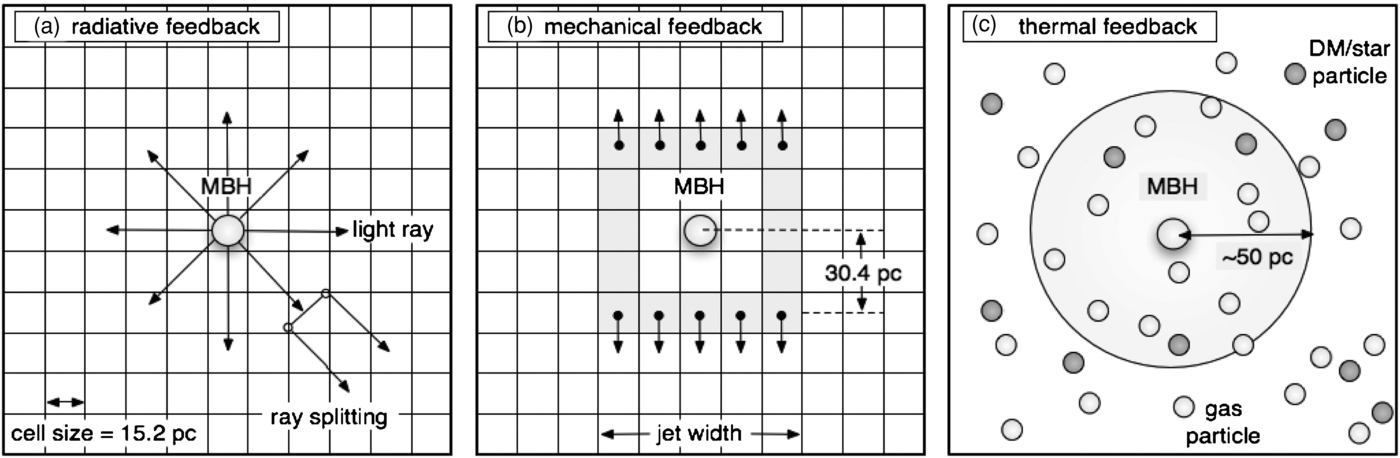}
\caption{\label{fig:method} Schematic views of numerical implementations of BH feedback. (a) Radiative feedback through ray tracing, (b) Mechanical jet feedback through momentum injection in  ``supercells'', and (c) thermal feedback that assumes some thermal conversion efficiency between radiative/mechanical feedback and  thermal energy.  Adopted from \citet{Kim11},  \textcopyright AAS. Reproduced with permission.}
\end{figure}

\paragraph{Numerical simulations} self-consistently model the large-scale structure, explicitly following the buildup of many halos from Gaussian density perturbations.  Their usual repertoire includes hydrodynamics, non-equilibrium gas cooling \citep{Grackle}, and formation and feedback models for stars and BHs \citep{Vogelsberger13}, with the most sophisticated simulations now including magnetic fields \citep{Pakmor13}, cosmic ray feedback \citep{Pfrommer17}, and radiative transfer \citep{Xu16_fesc, Pawlik17}.  They can be solved using an adaptive Cartesian grid, smooth particle hydrodynamics, a moving mesh, or meshless methods. Figure \ref{fig:method} shows an example of different methods of injecting (radiative, mechanical, and thermal) feedback energy from an accreting SMBH \citep{Kim11}. The biggest physics limitation is the limited spatial and mass resolution, necessitating ``sub-grid'' models of star/BH formation and feedback and any other type of physics occuring below this limit.  Particular attention must be paid toward energy balance, where any feedback should not be irradiated away before being converted into the kinetic gas energy through the hydrodynamical solver \citep{Katz92, Stinson06, Wise2012b}.  In the process, these models must be tuned to match observed properties of galaxies and their BHs, in which subsequently they can make predictions for future observations and determine the relative importance of each process, not unlike semi-analytical models.  A more logistical downside is the computational expense of these calculations, sometimes taking several months while using thousands of compute cores.  Nevertheless by directly simulating the system, these calculations can capture the complex and non-linear interplay between astrophysical processes, especially in the context of feedback loops.

\section{Initial growth of the First Black Holes}
\label{sec:initial}

The initial growth of the first BHs is highly dependent on its stellar progenitor and how it affected the adjacent environment, which can be encapsulated by the Bondi-Hoyle rate (\S~\ref{sec:accretion}). Afterward cosmological effects, such as mergers and associated gas mass accretion rates, play a significant role in the evolution of the seed BH and its host galaxy.  As the galaxy grows, both star formation and SMBH fueling will become self-regulated through their feedback processes \citep{Milosavljevic09, Park12}. In this section, we will review the gaseous properties surrounding the seed BH from a metal-free (Population III; Pop. III) star, a dense stellar cluster, and a direct-collapse BH.

\subsection{Population III Stellar Remnant Seeds}

\begin{figure}
\centering
\includegraphics[width=0.75\textwidth]{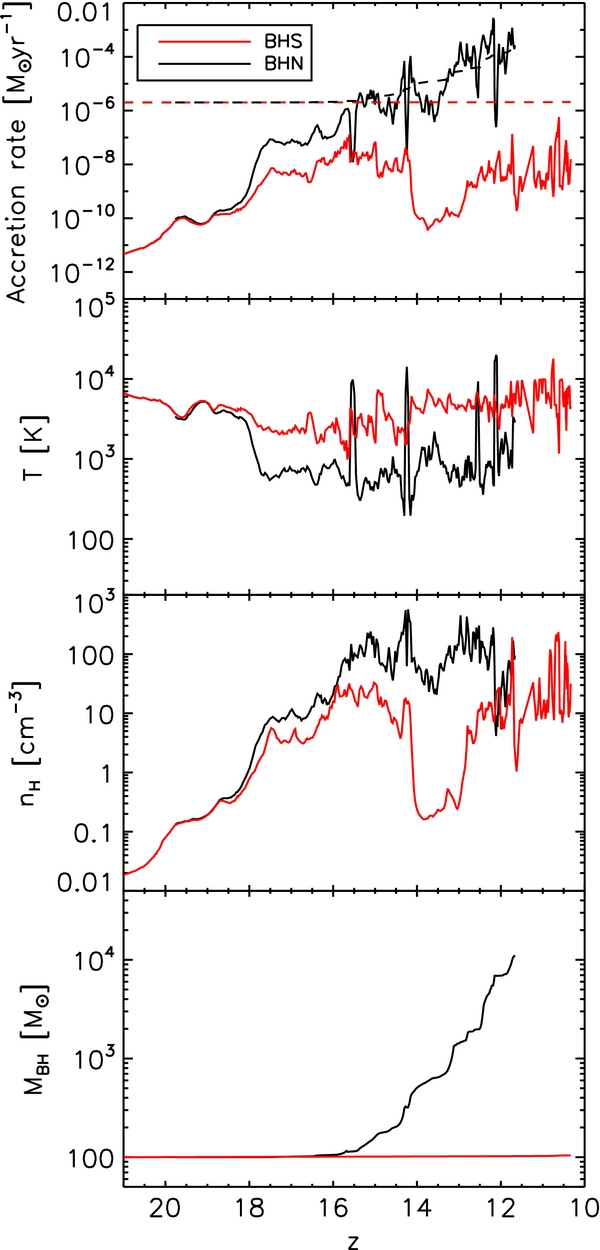}
\caption{\label{fig:pop3} Evolution of a Pop. III seed BH.  {\it Top to bottom:} BH accretion rate, temperature and gas number density of the immediate vicinity, and BH mass with (red) and without (black) radiative feedback.  The dashed line in the top panel marks the Eddington limit.  Adopted from \citet{Jeon12},  \textcopyright AAS. Reproduced with permission.}
\end{figure}

Observations of metal-poor stars and simulations of Pop. III formation (see also Chapter~4) have suggested that they have a top-heavy initial mass function \citep{Aoki2014, Susa14, Hirano2015}, resulting in a large production of BH remnants.  Furthermore being massive, they are prodigious producers of ionizing radiation, creating \hii{} regions on the kpc-scale.  Most of the gas in the host (mini-)halo with masses $\sim 10^6~\Ms$ is swept up by the $\sim 30~\kms$ shock associated with the ionization front, leaving behind a warm ($\sim 10^4 \unit{K}$) and diffuse ($\sim 1~\cubecm$) medium \citep{Kitayama2004, Whalen2004, Alvarez2006, Abel2007}.  The escape velocities of these halos are only a few \kms{}, so the halo is evacuated of nearly all of its gas after the shock leaves the halo \citep{Muratov13, Jeon14Recovery}.  This medium is not conducive for BH growth, and it must wait until a gas-rich merger that resupplies the dense gas reservoir.

Simulations following the growth of these seed BHs have been restricted to halos with a quiescent assembly history, only increasing by a factor of 10--100 to the atomic cooling limit of $\sim 10^8~\Ms$ over $\sim$200 million years.  However during this phase, X-ray radiative feedback greatly suppresses any further fueling, and the seed does not grow more than 1\% during this time at a typical rate of $\sim 0.01 \dot{M}_{\rm Edd}$ \citep{Alvarez2009, Jeon12}.  Figure \ref{fig:pop3} compares the accretion rate, BH mass, and temperature and density of the surrounding gas of a simulation with and without radiative feedback \cite{Jeon12}.  Accretion radiation allows the gas to maintain $T \sim 10^4 \unit{K}$ and lowers its density through thermal pressure forces.  If feedback is neglected, the BH grows by a factor of 100 at the Eddington rate when the halo undergoes a series of mergers at $z \simeq 14$ and approaches the atomic cooling limit at $z \simeq 12$, which is driven by a denser and cooler gas reservoir. \citet{Jeon2014} then considered Pop. III high mass X-ray binaries. After consuming the envelope of the secondary star, they have accretion rates up to eight orders of magnitude lower than the Eddington limit when hosted by minihalos, similar to isolated Pop. III stars.



Pop. III star formation relies on \hh{} cooling in minihalos, which can be dissociated by a Lyman-Werner background \citep{Dekel87,  machacek01}.  In this case, they will first form in larger minihalos with masses in the range $10^6 - 10^8~\Ms$ \citep{machacek01,Wise07A, OShea2008}.  Although there is more gas available, the star formation efficiencies are not much higher than lower mass minihalos, only totaling up to a few thousand solar masses
\citep{Johnson10, Xu16}.  Their total UV luminosity might not be sufficient to blow out the gas from the host halo.  This affected gas is then more likely to fall back into the halo center, aiding in early fueling of the BH seeds.  Some fraction of these stars will explode in supernova, perhaps triggering prompt star formation \citep{Whalen08}, and the gas reservoir would be consumed by further star formation and disrupted by their supernova.  However,
there has been little numerical focus on BH growth and feedback in these more massive minihalos.

\subsection{Intermediate Mass Black Hole Seeds}

Stellar collisions and dynamical instabilities can cause dense stellar clusters to form seed BHs with masses $\sim 10^3~\Ms$ \citep[][see also Chapter 7]{Rees1984, Devecchi2009, Davies2011ApJ, Alexander2014}.  Regardless of the formation mechanism, they all have massive stellar precursors that will photo-heat the immediate environment.  During the formation phase, dynamical effects between massive stars or BHs ($\sim 10~\Ms$) will drive dense gas inflows toward the center, fueling the seed BHs at super-Eddington rates \citep{Davies2011ApJ, Alexander2014}.  After this phase, further gas supply may be disrupted by stellar (radiative and supernova) and BH feedback.  If they are hosted in a minihalo, they might have a difficult time growing without external gas supplies from mergers, similar to the Pop. III seed BHs.  However, being embedded in a stellar cluster with $M_\star \simeq 10^4~\Ms$ \citep{Katz2015}, the deeper potential well could mitigate any fueling issues \citep{Park2016}.  If the host halo has a virial temperature $T_{\rm vir} \gsim 10^4 \unit{K}$ (i.e. an atomic cooler), then the seed BH should avoid these initial growing pains if it is centrally located in
the halo.  However, most of the attention has been on the formation of such intermediate-mass seed BHs, not the ensuing growth.

\subsection{Direct Collapse Black Hole Seeds}

Models of supermassive star (SMS) formation suggest that they form in metal-free atomic cooling halos that have little \hh{} content, limiting their cooling to $\gsim 8000 \unit{K}$ \citep[][see also Chapter~6]{Omukai2001,Oh2002}.  In addition, they must be fueled by rapid infalling gas flows, providing over $0.1~\Ms\unit{yr}^{-1}$ onto the SMS.  They can grow up to a mass of $10^6~\Ms$ with $\sim$10\% collapsing into a direct-collapse black hole (DCBH)
\citep{Begelman2006,  Begelman2008, Johnson2013b}, resulting in massive seed BHs in the range $M_{\rm BH} \simeq 10^4 - 10^5~\Ms$.

During the SMS main sequence, the stellar envelope is supported by radiation pressure with surface temperatures around 5000~K \citep[][see also Chapter~11]{Begelman12, Hosokawa12, Hosokawa2013, Schleicher13, Sakurai2015}. Thus their ionizing photon luminosities are extremely low $Q \sim 10^{44} - 10^{46} \unit{ph s}^{-1}$ and will have little effect on the surrounding medium that is rapidly falling toward it.  Also, a relativistic jet might be launched from the collapsing SMS powering a gamma-ray burst, disrupting any leftover accreting material \cite{Matsumoto15}.  In addition, some SMSs may have exploded in extremely energetic ($\sim 10^{53} - 10^{55} \unit{erg}$) supernovae \citep{Montero2012, Chen14}, totally disrupting the gaseous structures in the host halo \citep{Whalen2013}.

\begin{figure}[t]
\centering
\includegraphics[width=\textwidth]{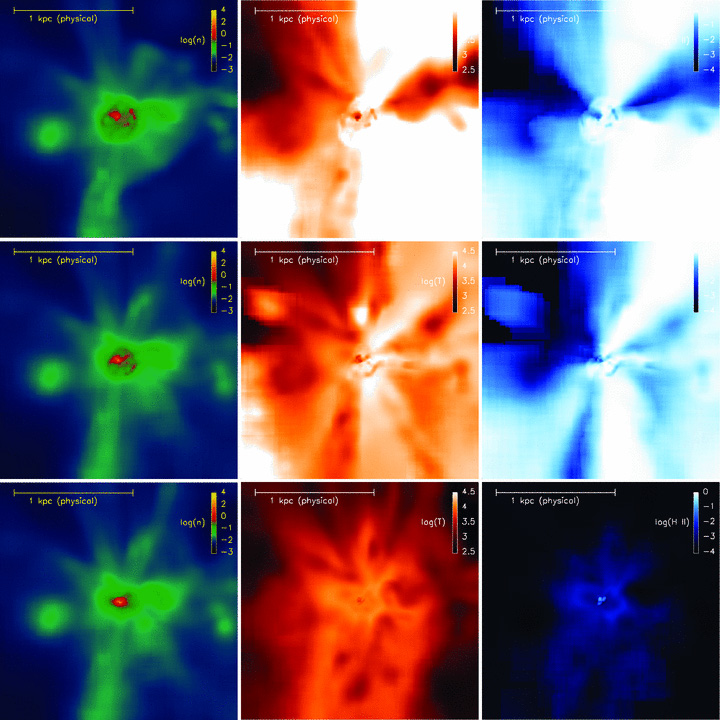}
\caption{\label{fig:dcbh} Radiative feedback from massive BH seeds with $M_{\rm BH} = 5 \times 10^4~\Ms$ (top), $2.5 \times 10^4~\Ms$  (middle), and $10^4~\Ms$ (bottom).  Projections of number density, temperature, and \hii{} fraction three million years after BH formation in a field of view of 2 proper kpc.  The more massive BH seed create larger \hii{} regions along with disrupting the nearby dense gas.  Adopted from \citet{Johnson10}, reproduced by permission of Oxford University Press / on behalf of the RAS.}
\end{figure}

Once the DCBH has fully formed, its feedback through radiation and jets should have a substantial impact on the host halo.  Because their masses are between $10^4$ and $10^5~\Ms$, most of their radiation will lie in the UV with a minor component in the X-rays \citep{Johnson2011, Pacucci2016}.  The UV radiation photo-heats the gas to $10^4 - 10^5 \unit{K}$ and drives down its density to 10--300~\cubecm{} after three million years \citep{Johnson2011}.  They
also found that the DCBH accretes near the Eddington limit immediately after formation but decreases to 1\% of the limit after a few million years.  Figure \ref{fig:dcbh} shows projections of the density, temperature, and ionized fraction of the host halo and its surroundings.  This radiation propagates through the surrounding metal-free medium and prompts Pop. III star formation through enhanced \hh{} formation \citep{Aykutalp17}.  A fraction of these metal-free
stars enriches the inner regions of the halo and partially disrupts the inflowing gas.  But the now metal-enriched gas will have a higher opacity in the X-rays.  These X-rays will photo-heat the gas to $T \simeq 10^6 - 10^7 \unit{K}$, driving outflows in the process \citep{Aykutalp14}.  The massive BH seed eventually settles into a self-regulated accretion pattern, having a duty cycle 5--50\%, depending on the gas supply, only growing by 10--20\% within 100
million years.  Similar to the less massive seed BHs, extended super-Eddington growth periods are probably dependent on cosmological mergers, bringing a fresh gas supply that could smother the central BH, creating an opacity-thick medium in and around the accretion disk that favors super-Eddington accretion \citep{Wyithe12, Pacucci2015, Kohei, Pezzulli16, Sakurai2016}.

\section{Incorporation into Larger Halos and Galaxies}
\label{sec:larger}

The observed $z \sim 6$ AGN are located in rare halos that have extremely rapid accretion histories, forming their first luminous objects at $z \sim 30$ and growing by a factor of $\sim 10^6$ with in 800 million years (see Figure \ref{fig:ps}).  They will be bombarded with infalling halos and gas, fueling intense star formation and central BH growth.  In the process, they will have several major mergers and hundreds, if not thousands, of minor mergers, all
containing the first generations of galaxies and most likely a population of seed BHs.  On the other hand, more typical galaxies, like the Milky Way, will grow at a more leisurely pace, whose progenitor halos can only support star formation after $z \sim 10$. But it is not a problem if their initial growth is stunted because they have several more billion years to grow into the observed AGN population at lower redshifts \citep{Miyaji15}.  Nevertheless, theories of SMBH growth during the early universe should capture both the typical and extreme cases.

At early times, a halo/galaxy is not guaranteed to have a central BH because of there is some probability that either a Pop. III star does not leave behind a BH remnant in the mass ranges $M_\star < 20~\Ms$ and $M_\star \simeq 140-260~\Ms$ \citep{Heger2003, Heger2010, Takahashi14} or a dense stellar cluster does not experience a runaway collapse. When the most massive progenitor does not host a central BH, the early galaxy will initially have a swarm of tens of stellar-mass BHs from Pop. III stars as their halos merge into the primary galaxy, moving at a velocity dispersion similar to the halo circular velocity \citep{Xu13}.  \citet{Habouzit17} found that only 20\% of galaxies with stellar masses $M_\star \sim 10^6~\Ms$ have central BHs in their simulations, and all galaxies with $M_\star > 10^8~\Ms$ have a central object.  For the remainder of this section, we will focus on early galaxies with central BHs.

Halo mergers can take a substantial fraction (tens of millions of years) of the Hubble time at $z > 10$.  During the merger, the BHs must sink into the potential well center through dynamical friction \citep{Chandrasekhar1943}, which takes even more time than the halo merger \citep{Tremmel17}.  For DCBH seeds, this is especially important because they are theorized to mainly form in satellite halos that are falling into an newly born galaxy.  They are thought to form $\sim 1 \unit{kpc}$ from the galaxy \citep{Dijkstra08,Agarwal14,chon16,Regan17} and then infall and migrate toward the galactic center, where conditions for further accretion are more favorable.

The tight time constraints for the $z \sim 6$ AGN suggests that SMBHs grow mainly through accretion than mergers \citep{Volonteri2005}.  BH mergers also have the complication of the merged remnant receiving a kick with $v \sim 300~\kms{}$, depending on the BH mass ratios and spins, which is greater than the vast majority of first generation of galaxies.  These kicked compact objects escape the galaxy and take several billion years to return to a galaxy \citep{Micic07, Whalen12}.

\begin{figure}[t]
\centering
\includegraphics[width=\textwidth]{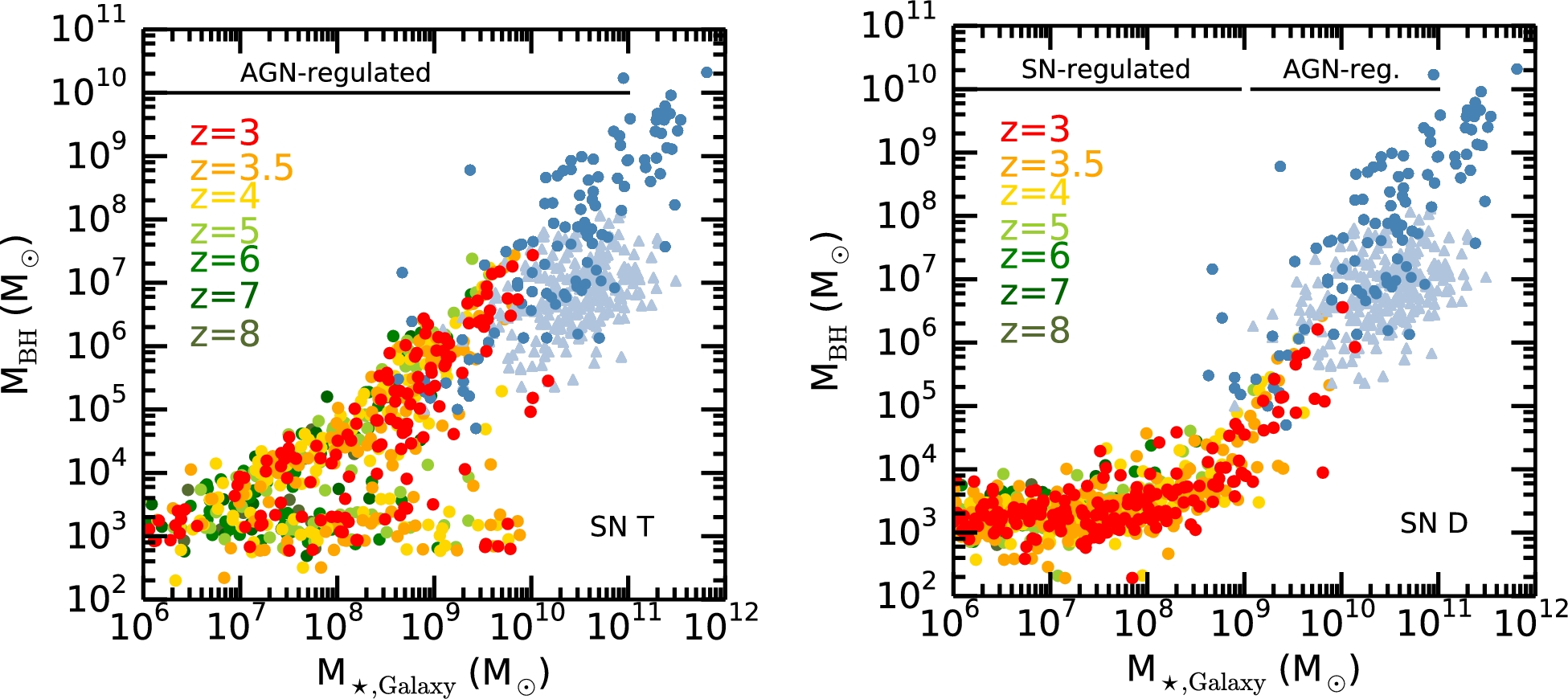}
\caption{\label{fig:galbh} Black hole mass as a function of stellar mass for a thermal supernova feedback (left) and delayed cooling supernova feedback (right) with the latter better matching observations.  The blue points show the observed present-day  relation \citep{Reines15}.  In galaxies with $M_\star \le 10^9~\Ms$, supernova feedback launches winds from the central regions, removing potential gas for black hole accretion.  Adopted from \citet{Habouzit17}, reproduced by permission of Oxford University Press / on behalf of the RAS.}
\end{figure}

High-redshift galaxies tend to have high specific star formation rates ($\textrm{sSFR} \equiv \dot{M}_\star/M_\star \simeq 1-50 \unit{Gyr}^{-1}$), doubling their stellar masses in as little as 20 million years and rarely having any organized rotation \citep{Wise2012, Ma17}.  However, there is a lack of low- and moderate-luminosity AGN at $z \gsim 6$ \citep{Willott11, Weigel15}.  Can the central BH keep pace with this furious growth
rate?  \citet{Volonteri17} used empirical scaling relations between low-mass galaxies and their central BHs to determine that such objects are more likely to have lower Eddington ratios ($\dot{M}_{\rm BH}/\dot{M}_{\rm Edd}$) than the observed luminous quasars.  This difference suggests that either moderate-luminosity AGN are intrinsically X-ray weak \citep{Luo14} or that they are heavily obscured by dense and dusty gas \citep{Pezzulli17}, which could be
especially true when gas inflow rates are high.  Using high-redshift galaxy formation simulations that captured the formation of Pop. III remnant BH seeds, \citet{Habouzit17} showed that SN feedback drove outflows that suppress BH accretion in low-mass ($M_\star < 10^9~\Ms$) galaxies, only growing by a factor of 10--100 up to $10^4~\Ms$ by $z = 3$, as depicted in Figure \ref{fig:galbh}.  However above this mass scale, the central BHs begin to grow faster at $\dot{M}_{\rm BH} \gsim 0.01 \dot{M}_{\rm Edd}$ for a substantial fraction of time.

Although the BH seeds and host galaxies do not lie on the \msigma{} relation, they should approach the observed relation at the present-day.  \citet{Volonteri09} explored how it evolves from high redshift to $z = 1$ when high-redshift progenitors were populated with ``light seeds'' and ``heavy seeds''.  The former have masses $\sim 100~\Ms$ corresponding to Pop. III remnants, and the latter exist in a mass range $10^2 - 10^6~\Ms$ that were derived from the \citet{Lodato2006} DCBH formation model.  The light seeds have a characteristic plume of ungrown seeds existing in galaxies with a velocity dispersion $\sigma = 50-100 \kms$ that then transition to the observed \msigma{} relation.  The heavy seeding model exhibits a BH mass floor at $10^5~\Ms$ in galaxies below this same velocity dispersion \citep{Agarwal13, Priya17}.



It is common practice for numerical simulations with a focus toward present-day galaxies and BHs to seed the smallest resolved halos ($M_{\rm vir} \sim 10^9 - 10^{10}~\Ms$) with a central BH with $M_{\rm  BH} \sim 10^5~\Ms$ that obeys the present-day \msigma{} relation \citep{Li07, DiMatteo2012, Dubois14, Taylor14, Illustris,  EAGLE, Feng14, Sijacki2015, Smidt17, Tenneti17}.  These seeding models assume that central BHs grow efficiently in halos below their mass resolution, however as previously discussed, they can be heavily affected by their own feedback and depletion of dense gas through supernova winds, if they host a central BH at all \citep{Aykutalp14, Habouzit17}.  Nevertheless, these large-scale galaxy simulations are accurate for galaxies similar in mass to the Milky Way and larger, but they may be lacking accuracy at high redshifts and low masses and cannot make predictions about such objects.  Because it is currently computationally prohibitive to directly follow the smallest star forming minihalos all the way to the most massive SMBHs, a good compromise in these large-scale simulations is to seed the smallest galaxies stochastically according to the stellar-BH mass ratios found in high-redshift, small-scale simulations or analytical models \citep{Bellovary11, Habouzit17} as a way forward before clever algorithms and more powerful supercomputers can model the entire formation sequence.

\section{Summary and Future Directions}
\label{sec:future}


Although there are still many open questions about the yet-to-be observed universe when the first generations of stars, galaxies, and black holes were abundant, we have a good understanding about the general progression of early galaxy formation and the growth of their (if any) central BHs.  One major uncertainty is the initial mass function of seed BHs (see also discussion in Chapter~9), which then affects their subsequent growth and feedback.  However it is clear that all of the progenitors have massive stars in common, suggesting that the initial growth phase will be slow in the warm and diffuse medium left behind.  Substantial growth will be delayed until gas-rich halo mergers provide additional fuel.

Better resolution and larger dynamic range are always an easy answer for future work, but we should continue to place emphasis on the importance of accurate sub-grid BH models \citep{Negri17} because, barring a breakthrough, a cosmological simulation that resolves and evolves the accretion disks and the associated radiation over the first billion years is not possible in at least the next decade.  Bondi-Hoyle accretion is idealized, whereas reality is nothing but ideal.  One could imagine having a low-resolution accelerated calculation existing inside the simulation, using the properties from the cosmological simulation as boundary conditions. For example, better models could include (1) radiation transport at many energies, (2) sub-grid models for a multi-phase accretion from the ISM to an accretion disk to the BH / jet \citep{Power11}, and (3) better estimating angular momentum transport and BH spin that would respectively influence the jet direction and radiative efficiency. Including these effects in sub-grid models could inform us when super-Eddington accretion occurs, if at all, in BHs evolving from their seeds to the monsters we observe at $z \sim 6$.

Future observatories, such as JWST, WFIRST, ATHENA, and thirty-meter class telescopes, will enlighten us with the spectral properties of a multitude of $z > 6$ galaxies and their AGN, which is the subject of the following chapters.  Mock observations of these objects \citep{Barrow17, Priya17, Volonteri17} are of critical importance to constrain BH growth scenarios and their relationship to host galaxies, using these upcoming spectacular photometric and spectral datasets of the first galaxies and black holes.

In the next chapter, the discussion on the growth of SMBHs will be complemented with specific scenarios that allow super-Eddington accretion and thus a more rapid growth. An account of the current observational status is given in Chapter~12. Predictions for gravitational wave observatories are outlined in Chapter~13, and future observational possibilities in Chapter~14.

\section*{Acknowledgments}
JHW is supported by National Science Foundation grants AST-1333360 and AST-1614333, NASA grant NNX17AG23G, and Hubble theory grants HST-AR-13895 and HST-AR-14326.  This research has made use of NASA's
Astrophysics Data System Bibliographic Services.


{
\bibliographystyle{ws-rv-har}    
\bibliography{ref}
}

\printindex[aindx]           
\printindex                  

\end{document}